\documentclass[10pt,conference]{IEEEtran}

\usepackage{amssymb}
\usepackage{amsxtra}
\usepackage{amsmath}
\usepackage{latexsym}
\usepackage{graphics}
\usepackage{verbatim}
\usepackage{epsfig}
\usepackage{setspace}
\usepackage{theorem}

\theorembodyfont{\normalfont\slshape}

\newtheorem{theorem}{Theorem}[section]
\newtheorem{lemma}[theorem]{Lemma}
\newtheorem{proposition}[theorem]{Proposition}
\newtheorem{corollary}[theorem]{Corollary}

\newtheorem{remark}[theorem]{Remark}

\def\S{\Sigma}

\def\Z{{\mathbb Z}}

\def\a{{\mathbf a}}

\def\c{{\mathbf c}}

\def\d{{\mathbf d}}

\def\w{{\mathbf w}}
\def\x{{\mathbf x}}
\def\y{{\mathbf y}}

\def\cB{{\mathcal B}}

\def\cF{{\mathcal F}}
\def\cG{{\mathcal G}}
\def\cH{{\mathcal H}}

\def\cP{{\mathcal P}}
\def\cS{{\mathcal S}}

\def\cV{{\mathcal V}}

\def\cX{{\mathcal X}}
\def\cY{{\mathcal Y}}

\def\per{{\text{per}}}

\begin{document}

\thispagestyle{empty}

\title{On the Period of a Periodic-Finite-Type Shift}

\author{\authorblockN{Akiko Manada and Navin Kashyap}
\authorblockA{Dept.\ Mathematics and Statistics\\
Queen's University\\
Kingston, ON, K7L 3N6, Canada.\\
Email: \texttt{\{akiko,nkashyap\}@mast.queensu.ca}}
}

\maketitle

\date{}
\maketitle

\renewcommand{\thefootnote}{$*$}
\footnotetext{This work was supported in part by a Discovery Grant 
from the Natural Sciences and Engineering Research Council (NSERC) of Canada.}

\renewcommand{\thefootnote}{\arabic{footnote}}
\setcounter{footnote}{0}

\begin{abstract} 
Periodic-finite-type shifts (PFT's) form a class of sofic shifts 
that strictly contains the class of shifts of finite type (SFT's). 
In this paper, we investigate how the notion of ``period'' inherent
in the definition of a PFT causes it to differ from an SFT, and 
how the period influences the properties of a PFT.
\end{abstract}


\section{Introduction\label{intro}}
Shifts of finite type (SFT's) are objects of fundamental importance 
in symbolic dynamics and the theory of constrained coding \cite{ML}.
A well-known example of an SFT would be the $(d,k)$ run-length 
limited ($(d,k)$-RLL) shift, where the number of 0's between 
successive 1's is at least $d$ and at most $k$.
Constrained codes based on these $(d,k)$-RLL shifts are used 
in most storage media such as magnetic tapes, CD's and DVD's.

A generalization of SFT's was introduced by Moision and Siegel \cite{MS2} 
who were interested in examining the properties of distance-enhancing 
constrained codes, in which the appearance of certain words is 
forbidden in a periodic manner. This new class of shifts, called 
periodic-finite-type shifts (PFT's), contains the class of SFT's and
some other interesting classes of shifts, such as constrained systems
with unconstrained positions \cite{BCF},\cite{SMNW}, and shifts arising
from the time-varying maximum transition run constraint \cite{PM}.
The class of PFT's is in turn properly contained within the class of 
sofic shifts \cite{MS}, a fact we discuss in more detail in 
Section~\ref{background}.

The properties of SFT's are now quite well understood (cf.\ \cite{ML}),
but the same cannot be said for PFT's. The study of PFT's has primarily 
focused on finding efficient algorithms for constructing their presentations 
\cite{BCF}, \cite{MS}, \cite{P1}. The difference between the definitions 
of SFT's and PFT's is quite small. An SFT is defined as a set of 
bi-infinite sequences (over some alphabet) that do not contain as subwords 
any word from a certain finite set. Thus, an SFT is defined by forbidding 
the appearance of finitely many words at any position of a 
bi-infinite sequence. A PFT is also defined by forbidding the 
appearance of finitely many words, except that these words are only 
forbidden to appear at positions of a bi-infinite sequence
that are indexed by certain pre-defined periodic integer sequences;
see Section~\ref{background} for a formal definition. This paper
aims to initiate a study of how the ``period'' inherent in the
definition of a PFT influences its properties.

After a review of relevant definitions and background in 
Section~\ref{background}, we will see in Section~\ref{period_influence}
that given an SFT $\cY$, we can associate with it a PFT $\cX$
in such a way that it is only the period that differentiates
$\cX$ from $\cY$. We then seek to understand how the period determines 
the properties of the PFT $\cX$ by means of a comparative study of $\cX$ 
and $\cY$. We investigate a different aspect of periods in 
Section~\ref{periods}, where we study the influence of the period of a PFT
$\cX$ on the periods of periodic sequences in $\cX$, and on the
periods of graphical presentations of $\cX$.

\section{Basic Background on SFT's and PFT's\label{background}}
We begin with a review of basic background, based on material from 
\cite{ML} and \cite{MS}. Let $\S$ be a finite set of symbols; 
we call $\S$ an \emph{alphabet}. We always assume that $|\S|=q\geq 2$
since $q=1$ gives us a trivial case. Let $\w=\ldots w_{-1}w_{0}w_{1}\ldots$ 
be a bi-infinite sequence over $\S$. A word (finite-length sequence) 
$u\in \S^n$ (for some integer $n$) is said to be a \emph{subword} of $\w$, 
denoted by $u\prec \w$, if $u=w_iw_{i+1}\ldots w_{i+n-1}$ 
for some integer $i$. If we want to emphasize the fact that 
$u$ is a subword of $\w$ starting at the index $i$, 
(\emph{i.e.}, $u=w_iw_{i+1}\ldots w_{i+n-1}$), we write $u\prec_i \w$. 
By convention, we assume that the empty word $\epsilon \in \S^0$ 
is a subword of any bi-infinite sequence. Also, we define $\sigma$ 
to be the shift map, that is, 
$\sigma(\w)=\ldots w^*_{-1}w^*_{0}w^*_{1}\ldots$ is the bi-infinite sequence 
satisfying $w^{*}_{i}=w_{i+1}$ for all $i$. 
 
Given a labeled directed graph $\cG$, where labels come from $\S$, 
let $S(\cG)$ be the set of bi-infinite sequences which are 
generated by reading off labels along bi-infinite paths in $\cG$. 
A \emph{sofic shift} $\cS$ is a set of bi-infinite sequences such that 
$\cS=S(\cG)$ for some labeled directed graph $\cG$. In this case, we say 
that $\cS$ is \emph{presented by} $\cG$, or that $\cG$ is a 
\emph{presentation} of $\cS$. It is well known that every sofic shift has 
a \emph{deterministic} presentation, \emph{i.e.}, a presentation such that 
outgoing edges from the same state (vertex) are labeled distinctly. 
For a sofic shift $\cS$, $\cB_n(\cS)$ denotes the set of words 
$u\in \S^n$ satisfying $u\prec \w$ for some bi-infinite sequence 
$\w$ in $\cS$, and $\cB(\cS)=\cup_{n\geq 0} \cB_n(\cS)$. 
A sofic shift $\cS$ is \emph{irreducible} if there is an irreducible 
(\emph{i.e.}, strongly connected) presentation of $\cS$, or equivalently, 
for every ordered pair of words $u$ and $v$ in $\cB(\cS)$, 
there exists a word $z\in \cB(\cS)$ such that $uzv \in \cB(\cS)$.   

A \emph{shift of finite type} (SFT) $\cY_{\cF'}$, with a finite set 
of forbidden words (a forbidden set) $\cF'$, is the set of all bi-infinite 
sequences $\w=\cdots w_{-1}w_{0}w_{1}\cdots$ over $\S$ such that
$\w$ contains no word $f' \in \cF'$ as a subword. That is, 
the finite number of words $f'$ in $\cF'$ are not in $\cB(\cY_{\cF'})$. 
A \emph{periodic-finite-type shift}, which we abbreviate as \emph{PFT},
is characterized by an ordered list of finite sets 
$\cF=(\cF^{(0)},\cF^{(1)}, \ldots, \cF^{(T-1)})$ and a \emph{period} $T$.
The PFT $\cX_{\{\cF,T\}}$ is defined as the set of all bi-infinite sequences 
$\w$ over $\S$ such that for some integer $r\in \{0,1,\ldots, T-1\}$, 
the $r$-shifted sequence $\sigma ^{r}(\w)$ of $\w$ satisfies 
$u\prec _i\sigma^{r}(\w)$ $\Longrightarrow $ 
$u\not \in \cF^{(i\ \mbox{mod}\ T)}$ for every integer $i$. For simplicity, 
we say that a word $f$ is in $\cF$ (symbolically, $f\in \cF$) if 
$f\in \cF^{(j)}$ for some $j$. Since the appearance of words 
$f\in \cF$ is forbidden in a periodic manner, note that $f$ can be in 
$\cB(\cX_{\{\cF,T\}})$. Also, observe that a PFT $\cX_{\{\cF,T\}}$ 
satisfying $\cF^{(0)}=\cF^{(1)}= \cdots =\cF^{(T-1)}$ is simply the 
SFT $\cY_{\cF'}$ with $\cF'=\cF^{(0)}$. Thus, SFT's are special cases 
of PFT's. We call a PFT \emph{proper} when it cannot be 
represented as an SFT. 

Any SFT can be considered to be an SFT in which every forbidden word 
has the same length. More precisely, given an SFT $\cY=\cY_{\cF^*}$, 
find the longest forbidden word in $\cF^*$ and say it has length $\ell$. 
Set $\cF'=\{f'\in \S^{\ell}: \mbox{$f'$ has some $f^*\in \cF^*$ 
as a prefix}\}$. Then, $\cY_{\cF^*}=\cY_{\cF'}$, and each word in 
$\cF'$ has the same length, $\ell$. Furthermore, we can also assume that
$\cB_{\ell}(\cY)=\S^{\ell}\setminus \cF'$ since if not (that is, if 
$\cB_{\ell}(\cY)\subsetneq \S^{\ell}\setminus \cF'$), 
every word in $(\S^{\ell}\setminus \cF')\setminus \cB_{\ell}(\cY)$ 
can be added to $\cF'$, without affecting $\cY$ in any way.   

Correspondingly, every PFT $\cX$ has a representation of the form 
$\cX_{\{\cF,T\}}$ such that $\cF^{(j)} = \emptyset$ for $1 \leq j \leq T-1$,
and every word in $\cF^{(0)}$ has the same length. An arbitrary representation
$\cX_{\{\cF,T\}}$ can be converted to one in the above form as follows.
If $f\in \cF^{(j)}$ for some $1\leq j\leq T-1$, list out 
all words with length $j+|f|$ whose suffix is $f$, add them to $\cF^{(0)}$, 
and delete $f$ from $\cF^{(j)}$. Continue this process until 
$\cF^{(1)}=\cdots=\cF^{(T-1)}=\emptyset $. Then, apply the method 
described above for SFT's to make every word in $\cF^{(0)}$ have 
the same length. 

It is known that PFT's belong to the class of sofic shifts.
\\[-18pt]

\begin{theorem}[Moision and Siegel, \cite{MS}]
All periodic-finite-type shifts $\cX$ are sofic shifts. 
That is, for any PFT $\cX$, there is a presentation $\cG$ of $\cX$.
\end{theorem}

Moision and Siegel proved the theorem by giving an algorithm that,
given a PFT $\cX$, generates a presentation, $\cG_\cX$, 
of $\cX$. We call the presentation $\cG_{\cX}$ the \emph{MS presentation} 
of $\cX$. The \emph{MS algorithm}, given a PFT $\cX$ as input, runs as follows.
\begin{enumerate}
\item Represent $\cX$ in the form $\cX_{\{\cF,T\}}$, such that 
every word in $\cF$ has the same length $\ell$ and belongs to $\cF^{(0)}$.
\item Prepare $T$ copies of $\S^{\ell}$ and name them 
$\cV^{(0)}, \cV^{(1)}, \ldots, \cV^{(T-1)}$.
\item Consider the words in $\cV^{(0)}, \cV^{(1)}, \ldots, \cV^{(T-1)}$ 
as states. Draw an edge labeled $a\in \S$ from 
$u=u_1u_2\cdots u_{\ell}\in \cV^{(j)}$ to 
$v=v_1v_2\cdots v_{\ell}\in \cV^{(j+1 \mod{T})}$ 
if and only if $u_2\cdots u_{\ell}=v_1\cdots v_{\ell-1}$ and $v_{\ell}=a$.
\item Remove states corresponding to words in $\cF^{(0)}$ from $\cV^{(0)}$, 
together with their incoming and outgoing edges. Call this labeled 
directed graph $\cG'$. 
\item If there is a state in $\cG'$ having only incoming edges or 
only outgoing edges, remove the state from $\cG'$ as well as its 
incoming or outgoing edges. Continue this process until we cannot 
find such a state. The resulting graph $\cG_{\cX}$ is a presentation of $\cX$. 
\end{enumerate}

\begin{remark}
It is evident that the MS presentation of a PFT is always deterministic. 
Also, for a path $\alpha $ in $\cG_{\cX}$ with length $|\alpha|\geq \ell$, 
$\alpha$ terminates at some state that is a copy of $u=u_1u_2\ldots u_{\ell}$ 
iff the length-$\ell$ suffix of the word generated by $\alpha$ 
is equal to $u$.
\label{terminal_rem}
\end{remark}

\section{Influence of the Period $T$ on a PFT\label{period_influence}}
From this point on, whenever we consider an SFT $\cY_{\cF'}$ in this paper, 
we will implicitly assume that every forbidden word in $\cF'$ has the 
same length $\ell$, and that $\cB_{\ell}(\cY)=\S^{\ell}\setminus \cF'$. 
As we observed in the previous section, there is no loss of generality
in doing so. Given an SFT $\cY_{\cF'}$, consider the PFT 
$\cX =\cX_{\{\cF,T\}}$ in which 
$$\cF=(\cF^{(0)}, \cF^{(1)}, \ldots, \cF^{(T-1)})=
(\cF',\emptyset ,\ldots, \emptyset ).$$
While $\cY_{\cF'} \subseteq \cX_{\{\cF,T\}}$, equality
does not hold in general. Note that it is only the influence of the 
period $T$ that causes the shifts $\cX = \cX_{\{\cF,T\}}$ and 
$\cY = \cY_{\cF'}$ to differ. So, a comparative study of 
$\cX$ and $\cY$ is a useful means of understanding how the period $T$
determines the properties of the PFT $\cX$. In this section, 
we present a sampling of results that illustrate how properties 
of the SFT $\cY$ can affect those of the PFT $\cX$. 

The following result, which shows that the irreducibility of $\cY$ 
has a significant effect on the irreducibility of $\cX$, 
may be considered typical of the comparative study proposed above.
\\[-18pt]

\begin{theorem}
Suppose that $\cY=\cY_{\cF'}$ is an irreducible SFT.
Let $\cX=\cX_{\{\cF, T\}}$ be the PFT satisfying
$$\cF=(\cF^{(0)}, \cF^{(1)}, \ldots, \cF^{(T-1)})=
(\cF',\emptyset , \ldots,\emptyset ).$$
If there exists a periodic bi-infinite sequence $\y$ in $\cY$ with 
a period $p$ satisfying $p \equiv 1 \pmod{T}$, 
then the MS presentation, $\cG_{\cX}$, of $\cX$ is irreducible as a graph.
That is, $\cX$ is irreducible. 
\label{irred.thm}
\end{theorem}
\mbox{}\\[-4ex]
\emph{Proof\/}: Throughout this proof, for a path $\eta$ in a graph, 
let $s(\eta)$ and $t(\eta)$ be the starting state and the terminal state,
respectively, of $\eta$ in the graph. Also, for a state 
$v=v_1v_2\ldots v_{\ell}$ in $\cG_{\cX}$, $v\in \cV^{(j)}$ is denoted by 
$v^{(j)}$ for $0\leq j\leq T-1$.

Let $\cG'$ be the graph defined in Step 4 of the MS algorithm. 
Consider the subgraph $\cH$ of $\cG'$ that is induced by the states 
in $\S^{\ell}\setminus \cF'$. Since $\S^{\ell}\setminus \cF'=\cB_{\ell}(\cY)$, 
all states in $\cH$ have incoming edges and outgoing edges. 
Hence, $\cH$ is a subgraph of $\cG_{\cX}$.

Key points of the proof are the following.\\
\emph{Claim 1\/}: $\cH$ is a presentation of $\cY$.\\
\emph{Claim 2\/}: $\cH$ is irreducible as a graph if 
there exists a periodic bi-infinite sequence $\y$ in $\cY$ with a period 
$p$ satisfying $p \equiv 1 \pmod{T}$.

Once these claims are proved, it is straightforward to check that 
the MS presentation $\cG_{\cX}$ of $\cX$ is irreducible.
Note that the graph $\cG'$ is obtained from $\cH$ by 
adding words in $\cF^{(0)}$ to $\cV^{(1)}, \cV^{(2)}, \ldots, \cV^{(T-1)}$ 
and corresponding incoming and outgoing edges. 
Observe that (by Step 5 of the MS algorithm) a word $f' \in \cF^{(0)}$ 
is a state in $\cG_{\cX}$ if and only if there exist paths 
$\rho _1$, $\rho _2$ in $\cG'$ satisfying 
$s(\rho _1)=f'$, $t(\rho _1)\in \S^{\ell} \setminus \cF'$ and 
$s(\rho _2)\in \S^{\ell}\setminus \cF'$, $t(\rho _2)=f'$. 
Since $\cH$ is irreducible, $\cG_{\cX}$ is irreducible as well.\\[4pt]
\underline{\emph{Proof of Claim 1\/}}:
We need to show that $S(\cH)\subseteq \cY$ and $\cY \subseteq S(\cH)$. 
It is clear that $S(\cH)\subseteq  \cY$ since, by Remark~\ref{terminal_rem}, 
there is no path in $\cH$ which generates words in $\cF'$.

Conversely, take an arbitrary bi-infinite sequence 
$\x=\ldots x_{-1}x_0x_1\ldots \in \cY$. Since $f'\not\prec \x$ 
for every forbidden word $f'\in \cF'$, we see that for any integer $i$, 
the states corresponding to $x_{i-\ell+1}x_{i-\ell+2}\ldots x_{i}$ 
are in $\cH$. Therefore, there exists an edge labeled $x_{i+1}$ from 
$x_{i-\ell+1}x_{i-\ell+2}\ldots x_{i}\in\cV^{(j)}$ to 
$x_{i-\ell+2}\ldots x_{i}x_{i+1}\in \cV^{(j+1 \mod T)}$ 
for all integers $i$ and $0\leq j\leq T-1$.    
Hence, $\x\in S(\cH)$, that is, $\cY \subseteq S(\cH)$. \\[4pt]
\underline{\emph{Proof of Claim 2\/}}:
A periodic bi-infinite sequence $\y\in \cY$ with period 
$p \equiv 1 \pmod{T}$ can be written as $\y=(y_1y_2\ldots y_n)^{\infty}$,
for some $y_1y_2\ldots y_n\in \S^n$, where $n$ is some multiple of $p$ 
satisfying $n \equiv 1 \pmod{T}$ and $n\geq \ell$. 

As $\y\in \cY$, $y_{n-\ell+1}\ldots y_ny_1y_2\ldots y_n \in \cB(\cY)$. 
Thus, for every $i \in \{0,1,\ldots, T-1\}$, 
there exists a path $\alpha$ in $\cH$ satisfying 
$s(\alpha)=z^{(i)}=y_{n-\ell+1}\ldots y_n$ and generating $y_1y_2\ldots y_n$. 
Observe that $t(\alpha)$ is also $z^{(i')}=y_{n-\ell+1}\ldots y_n$  
for some $i' \in \{0,1,\ldots, T-1\}$. 
However, since $|y_1y_2\ldots y_n| = n\equiv 1 \pmod{T}$, 
we have $i' = i+1 \mod{T}$. 
This automatically implies that for the word $z=y_{n-\ell+1}\ldots y_n$ 
in $\cB(\cY)$, there is a path $\beta_{jk}$ in $\cH$ 
such that $s(\beta_{jk})=z^{(j)}$ and $t(\beta_{jk})=z^{(k)}$ for any 
ordered pair $(j,k)$, where $0\leq j,k\leq T-1$. 
 
Now take an arbitrary pair of states $u^{(r)}$ and $v^{(s)}$ in $\cH$. 
Since $\cY$ is irreducible, there exist words $w'$ and $w^{*}$ in $\cB(\cY)$ 
so that $uw'z$ and $zw^{*}v$ are in $\cB(\cY)$.  
Thus, there exists a path $\gamma $ generating $w'z$ such that 
$s(\gamma )=u^{(r)}$ and $t(\gamma )=z^{(j)}$ for some $0\leq j\leq T-1$, 
and a path $\delta $ generating $w^{*}v$ such that
$s(\delta )=z^{(k)}$ for some $0\leq k\leq T-1$ and $t(\delta )=v^{(s)}$.  
As there is a path $\beta_{jk}$ from $z^{(j)}$ to $z^{(k)}$ from the 
argument above, we have a path $\gamma\beta_{jk}\delta$ starting from 
$u^{(r)}$ and terminating at $v^{(s)}$. 
Hence, the presentation $\cH$ is irreducible as a graph.
\endproof \mbox{} \\[-20pt]


From Theorem~\ref{irred.thm}, we can obtain the following corollary.
\\[-18pt]
\begin{corollary}
Let $\cY=\cY_{\cF'}$ be an irreducible SFT such that $|\cF'|<|\S|$. 
Then for all $T\geq 1$, the PFT $\cX=\cX_{\{\cF, T\}}$ with 
$$\cF=(\cF^{(0)}, \cF^{(1)}, \ldots, \cF^{(T-1)})=
(\cF',\emptyset , \ldots,\emptyset )$$
is irreducible. 
\label{uni-for.cor}
\end{corollary}
\mbox{}\\[-4ex]
\emph{Proof\/}: 
Since $|\cF'|<|\S|$, 
there is a symbol $a\in \S$ which is not used as the first symbol of any word in $\cF'$. 
Hence, the bi-infinite sequence $\a=a^{\infty}$ is in $\cY$. 
As $\a$ has period $1$, we have from Theorem~\ref{irred.thm} that 
$\cX$ is irreducible.
\endproof \mbox{}\\[-20pt]

The proof of Theorem~\ref{irred.thm} shows that the SFT $\cY=\cY_{\cF'}$ 
has a presentation $\cH$ that is a subgraph of the MS presentation 
$\cG_{\cX}$ of $\cX=\cX_{\{\cF,T\}}$, where 
$\cF=(\cF',\emptyset , \ldots,\emptyset )$. 
This fact may allow us to compare some of the invariants
associated with the two shifts $\cY$ and $\cX$, for example, 
their entropies and their zeta functions (see \cite[Chapters~4 and 6]{ML}). 
The entropy (or the Shannon capacity) $h(\cS)$ of a sofic shift $\cS$ 
can be computed from a deterministic presentation $\cG$ of $\cS$ as 
follows: $h(\cS)=\log_2\lambda$, where $\lambda$ is the largest 
eigenvalue of the adjacency matrix $A_{\cG}$ of $\cG$. Equivalently, 
$\lambda$ is the largest root of the characteristic polynomial 
$\chi_{A_{\cG}}(t)=\det(tI-A_{\cG})$ of $A_{\cG}$ 
(see, \emph{e.g.}, \cite[Chapter~4]{ML}).  

Returning to the shifts $\cX$ and $\cY$ as above, since $\cH$ is a 
subgraph of $\cG_{\cX}$, it may be possible to express the 
characteristic polynomial of $A_{\cG_{\cX}}$ in terms of the 
characteristic polynomial of $A_{\cH}$. This would allow us to compare 
the entropies of $\cX$ and $\cY$. However, this seems to be hard to do 
in general. We have a partial result in the special case when 
$\cY=\cY_{\cF'}$ with $|\cF'|=1$, and $\cX=\cX_{\{\cF,2\}}$,
as we describe next.   

Recall that $|\S|=q$. Now suppose that $\cY=\cY_{\cF'}$ is an SFT with 
the set $\cF'$ consisting of a single forbidden word $f'$, 
and $\cX=\cX_{\{\cF, 2\}}$ is the PFT with period 2 and 
$\cF=(\cF^{(0)}, \cF^{(1)})=(\{f'\},\emptyset)$. Also, 
let $A_{\cG_{\cX}}$ be the adjacency matrix of the MS presentation 
$\cG_{\cX}$ of $\cX$, and let $A_{\cH}$ be that of the subgraph $\cH$ 
of $\cG_{\cX}$ induced by the states in $\S^{\ell}\setminus \{f'\}$. 
Observe that the matrix $A_{\cG_{\cX}}$ is a 
$(2q^{\ell}-1)\times (2q^{\ell}-1)$ 0-1 matrix.  
Without loss of generality, for $A_{\cG_{\cX}}$, we can assume the 
following. 
\begin{itemize}
\item The first $q^{\ell}-1$ rows and columns correspond to states in 
$\cV^{(0)}$, and the last $q^{\ell}$ rows and columns correspond to those 
in $\cV^{(1)}$. 
\item Assign $f'\in \cV^{(1)}$ to the $(2q^{\ell}-1)$-th row and column, 
and arrange the first row so that the $(1,2q^{\ell}-1)$-th entry of 
$A_{\cG_{\cX}}$ is 1. 
\item Let $u\in \cV^{(1)}$ be such that the longest proper suffix of $u$ 
is equal to that of $f'$. Assign this $u$ to the $q^{\ell}$-th row and 
column so that the $q^{\ell}$-th row and the $(2q^{\ell}-1)$-th row 
are the same.  
\end{itemize} 
For a matrix $M$, set $M^{(i,j)}$ to be the submatrix of $M$ obtained 
by deleting its $i$-th row and $j$-th column.
Then, observe that $A_{\cG_{\cX}}^{(2q^{\ell}-1,2q^{\ell}-1)}=A_{\cH}$.  
In this case, by applying elementary row operations to the matrix 
$N=tI-A_{\cG_{\cX}}$, we have  
\begin{equation} 
\chi_{A_{\cG_{\cX}}}(t)=\det(N)=
\begin{vmatrix}
B & \c \\
\d & t
\end{vmatrix},
\end{equation}
%
where $B$ is a $(2q^{\ell}-2)\times (2q^{\ell}-2)$ matrix satisfying 
$\det(B)=\chi_{A_{\cH}}(t)$, $\c$ is the $(2q^\ell-2) \times 1$ 
column vector $[-1\ 0\ \ldots\ 0]^T$, and $\d \in \{-1,0\}^{2q^\ell-2}$. 
Using the form given in $(1)$ for $\det(N)$, we can derive the 
following theorem. The complete proof will be
published in the full version of this paper. 

\begin{theorem}
Let $\cY=\cY_{\cF'}$ and $\cX=\cX_{\{\cF,2\}}$ be the SFT and PFT 
described above, respectively. Then, the characteristic polynomial 
$\chi _{A_{\cG_{\cX}}}(t)$ of the adjacency matrix $A_{\cG_{\cX}}$ 
is given by 
$$
\chi _{A_{\cG_{\cX}}}(t) = 
t (\chi_{A_{\cH}}(t)+(-1)^{q^{\ell}}\det(B^{(1,q^{\ell})})).
$$
\label{char_poly_thm}
\end{theorem} 
\mbox{} \\[-40pt]

\section{Periods in PFT's\label{periods}}

The period $T$ involved in the description of a PFT is not the only 
notion of ``period'' that can be associated with the shift. For any shift
$\cX$, we can always define its \emph{sequential period}, $T_{seq}^{(\cX)}$,
to be the smallest period of any periodic bi-infinite sequence in $\cX$.
Furthermore, if $\cX$ is an irreducible sofic shift, 
we can define a ``graphical period'' for it as follows. 
Let $\cG$ be a presentation of $\cX$ with 
state set $\cV(\cG)=\{V_1, \ldots, V_r\}$. For each $V_i \in \cV(\cG)$,
define $\per(V_i)$ to be the greatest common divisor (gcd) of the 
lengths of paths (cycles) in $\cG$ that begin and end at $V_i$,
and further define $\per(\cG) = \gcd(\per(V_1),\ldots,\per(V_r))$.
It is well known that when $\cG$ is irreducible, $\per(V_i) = \per(V_j)$
for each pair of states $V_i,V_j \in \cV(\cG)$, 
and hence $\per(\cG) = \per(V)$ for any $V \in \cV(\cG)$. 
The \emph{graphical period}, $T_{graph}^{(\cX)}$, of an irreducible 
sofic shift $\cX$ is defined to be the least $\per(\cG)$ 
of any irreducible
presentation $\cG$ of $\cX$. 

Given a PFT $\cX$, define its \emph{descriptive period}, $T_{desc}^{(\cX)}$,  
to be the smallest integer among all $T^*$ such that 
$\cX=\cX_{\{\cF^*,T^*\}}$ for some $\cF^*$. In this section, we determine
what influence, if any, the descriptive period of a PFT has on its 
sequential and graphical periods.

 
Let $\cX=\cX_{\{\cF,T\}}$ be an irreducible PFT, and let $\cG$ 
be an irreducible presentation of $\cX$. Proposition 1 of \cite{MS} 
says that if $\cX$ is proper, then $\gcd(\per(\cG),T)\not=1$. 
Using that proposition, we can obtain the following result, which
shows that a proper PFT $\cX$ can have $T_{desc}^{(\cX)}$ arbitrarily 
larger than $T_{seq}^{(\cX)}$. \\[-18pt]

\begin{proposition}
Suppose that $\cY=\cY_{\cF'}$ is an irreducible SFT, such that the
bi-infinite sequence $a^{\infty}\in \cY$ for some $a \in \S$. 
Let $\cX=\cX_{\{\cF, T\}}$ be the PFT satisfying  
$$\cF=(\cF^{(0)}, \cF^{(1)}, \ldots, \cF^{(T-1)})=
(\cF',\emptyset , \ldots,\emptyset ).$$
Then, $a^{\infty} \in \cX$, so $T_{seq}^{(\cX)}=1$. Furthermore, 
if $\cX$ is a proper PFT and $T$ is prime, we have $T_{desc}^{(\cX)}=T$. 
\label{desc_pro}
\end{proposition}
\mbox{}\\[-4ex]
\emph{Proof\/}: Since $\cY \subseteq \cX$, it is clear that 
$a^{\infty} \in \cX$, and hence, $T_{seq}^{(\cX)}=1$. 
Now, let $\cX = \cX_{\{\cF, T\}}$ be a proper PFT with $T$ prime.
First observe that the MS presentation $\cG_{\cX}$ of $\cX$ is irreducible 
since the bi-infinite sequence $\a=a^{\infty }$ is in $\cY$ and $\a$ 
has period 1. Also, note that $\per(\cG_{\cX})$ must be $kT$ for 
some $k\geq 1$ from the construction of $\cG_{\cX}$. 
However, if we consider the period of the states $a^{\ell}$ in $\cG_{\cX}$, 
it is $T$. Thus, $\per(\cG_{\cX})=T$ by the irreducibility of $\cG_{\cX}$. 
Since $\cX$ is proper, we have from Proposition 1 of \cite{MS} 
that $\gcd(\per(\cG_{\cX}), T^*)\not=1$ for all $T^*$ satisfying  
$\cX=\cX_{\{\cF^*, T^* \}}$. As $T$ is prime, 
$\gcd(\per(\cG_{\cX}), T')=\gcd(T, T')=1$ for all $T'<T$. 
Therefore, $T$ is the descriptive period of $\cX$.
\endproof\mbox{} \\[-20pt]

For example, consider an SFT $\cY=\cY_{\cF'}$ with a 
forbidden set $\cF'=\{b^2\}$ for some $b\in \S$.
Then, $\cY$ is irreducible, and $a^{\infty}\in \cY$ 
for any $a\in \S\setminus \{b\}$. In this case, for a 
PFT $\cX=\cX_{\{\cF,T\}}$ with $T$ prime,
such that $\cF=(\{b^2\},\emptyset , \ldots,\emptyset )$, 
it may be verified that $\cX$ is proper, and hence, $T=T_{desc}^{(\cX)}$. 

Conversely, $T_{seq}^{(\cX)}$ can be arbitrarily larger 
than $T_{desc}^{(\cX)}$ for proper PFT's $\cX$. 
We present such an example next.
 
Set $\S=\{0,1\}$. We define a sliding-block map $\psi$ as follows:
for a non-empty word $u=u_1u_2\ldots u_n\in \S^n$,  
(resp.\ a bi-infinite sequence $\w=\ldots w_{-1}w_0w_1\ldots$ over $\S$), 
define $\psi(u)=u^{*}_1u^{*}_2 \ldots u^{*}_{n-1}$, 
where $u^{*}_i=u_i+u_{i+1}\! \pmod{2}$ for $1\leq i \leq n-1$ 
(resp.\ $\psi(\w)=\ldots w^*_{-1}w^*_0w^*_1\ldots$, 
where $w^{*}_i=w_i+w_{i+1}\! \pmod{2}$ for each $i$). 
By convention, $\psi(u)=\epsilon $ when $u\in \S^1$.
For $k\geq 1$, consider the PFT $\cX_k=\cX_{\{\cF_{k}, 2\}}$ 
with $\cF_k=(\cF_{k}^{(0)},\cF_{k}^{(1)})$, defined as follows.
\begin{itemize}
\item $\cF_k^{(1)}=\emptyset $ for all $k\geq 1$.
\item $\cF_1^{(0)}=\{0\}$, and for $k\geq 2$, we set 
  $\cF_{k}^{(0)}=\psi^{-1}(\cF_{k-1}^{(0)})$.
That is, $\cF_{k}^{(0)}$ is the inverse image of 
$\cF_{k-1}^{(0)}$ under $\psi$. 
\end{itemize}

It is easy to see that for each $k\geq 1$, every word 
$f \in \cF_k^{(0)}$ has length $|f|=k$, and in particular, 
we have $0^k \in \cF_k^{(0)}$. Moreover, as $\psi$ is a 
two-to-one mapping, we have $|\cF_k^{(0)}|=2^{k-1}$. 
The following proposition contains another useful observation 
concerning $\psi$. We omit the straightforward proof by induction. 

\begin{proposition}
For a binary word $u = u_1 u_2 \ldots u_r$ of length $r > m$, let
$u_1^*u_2^*\ldots u_{r-m}^* = \psi^m(u)$. If $m = 2^j$ for some 
$j \geq 0$, then $u^*_i=u_{i}+u_{i+2^j} \! \pmod{2}$ for $1\leq i\leq r-m$. 
Furthermore, if $m = 2^j - 1$ for some $j \geq 0$, then 
$u^*_i=u_{i}+u_{i+1}+\cdots +u_{i+2^j-1} \! \pmod{2}$ 
for $1\leq i\leq r-m$.
\label{forbid_pro}
\end{proposition} 

The corollary below simply follows from the fact that for any 
$f \in \cF_k^{(0)}$, we must have $\psi^{k-1}(f) = 0$. \\[-18pt]

\begin{corollary}
If $z \in \S^{2^j}$ (for some $j \geq 0$) has an odd number of 1's,
then $z \notin \cF_{2^j}^{(0)}$.
\label{odd1_cor}
\end{corollary}

We next record some important facts about the PFT's $\cX_k$. \\[-18pt]

\begin{proposition} 
For $k \geq 1$, the following statements hold:
(a)\ $\cX_{k+1}=\psi^{-1}(\cX_{k})$; 
(b)\ $\cX_k$ is irreducible iff $1\leq k\leq 6$; and
(c)\ $\cX_k$ is a proper PFT.
\label{Xk_prop}
\end{proposition}
\mbox{}\\[-4ex]
\emph{Proof\/}:
Statement (a) follows straightforwardly from the definition of the PFT's 
$\cX_k$. 

For (b), first note that $\cX_k$ is irreducible for $1\leq k\leq 6$ 
since its MS presentation may be verified to be irreducible as a graph. 
When $k=7$, it can be shown that $\cX_k$ is not irreducible, 
which implies that $\cX_k$ is not irreducible when $k\geq 7$ by (a).

To prove (c), suppose to the contrary that $\cX_k$ is not a proper PFT 
for some $k \geq 1$. Then, $\cX_k=\cY$ for some SFT $\cY = \cY_{\cF'}$, 
where every forbidden word in $\cF'$ has the same length, $\ell$.
Pick a $j \geq 0$ such that $2^j \geq k$, and set $r=2^j-k$.
By (a) above, $\cX_{2^j} = \psi^{-r}(\cX_k) = \psi^{-r}(\cY)$.
Note that $\psi^{-r}(\cY)$ is also an SFT, with forbidden set 
$\psi^{-r}(\cF')$. All words in $\psi^{-r}(\cF')$ have
length $\ell' = \ell + r$.

For the PFT $\cX_{2^j}$, observe that the bi-infinite sequence 
$\w=(0^{2^j-1}1)^{\infty}0^{2^j}(10^{2^j-1})^{\infty}$ is in $\cX_{2^j}$ 
as $\w$ contains a word in $\cF^{(0)}_{2^j}$ (\emph{i.e.}, $0^{2^j}$) 
only once, by Corollary~\ref{odd1_cor}. Therefore, 
every subword of $\w$ is in $\cB(\cX_{2^j}) = \cB(\psi^{-r}(\cY))$.  

Now, consider the bi-infinite sequence
$$\w'=(0^{2^j-1}1)^{\infty}0^{2^j} (10^{2^j-1})^{2\ell'+1}1 
0^{2^j} (10^{2^j-1})^{\infty}.$$
Note that every length-$\ell'$ subword of $\w'$ is also
a subword of $\w$, and hence, is in $\cB(\psi^{-r}(\cY))$.
This implies that $\w' \in \psi^{-r}(\cY)$. 
For the two distinct indices $m,n$ $(m<n)$ such that  
$0^{2^j} \prec_m \w'$ and $0^{2^j} \prec_n \w'$, 
we have $n-m= 2^j(2\ell'+2)+1$, so that $m \not\equiv n\! \pmod{2}$. 
But, since $0^{2^j} \in \cF_{2^j}^{(0)}$, this implies that
$\w' \not \in \cX_{2^j}$, which is a contradiction.  
\endproof \mbox{}\\[-20pt]

Statement (c) of Proposition~\ref{Xk_prop} implies that 
$T_{desc}^{(\cX_k)} = 2$ for all $k \geq 1$. In contrast, the 
following theorem shows that $T_{seq}^{(\cX_k)}$ grows arbitrarily 
large as $k \rightarrow \infty$.  \\[-12pt]
 
\begin{theorem}
For any $j\geq 0$ and $2^j+1\leq k\leq 2^{j+1}$, the periods of
periodic sequences in $\cX_k$ must be multiples of $2^{j+1}$. 
\label{mult.thm}
\end{theorem}

To prove Theorem~\ref{mult.thm}, we need the next three lemmas. We omit
the simple proof of the first lemma. \\[-18pt]

\begin{lemma}
If $\x \in \{0,1\}^\Z$ is a periodic sequence, then so is $\psi(\x)$.
Furthermore, any period of $\x$ is also a period of $\psi(\x)$.
\label{periodic_lem}
\end{lemma}


\begin{lemma}
For any $j\geq 0$, $\cF^{(0)}_{2^j+1}=\{f^*f^*_1 : 
f^*=f^*_1f^*_2\ldots f^*_{2^j}\in \S^{2^j}\}$.
\label{forbidden_lem}
\end{lemma}
\mbox{}\\[-4ex]
\emph{Proof\/}:
Recall that for a word $f\in \cF^{(0)}_{2^j+1}$, $\psi^{2^j}(f)=0$. 
Since Proposition~\ref{forbid_pro} shows that 
$\psi^{2^j}(f)=f_{1}+f_{2^j+1} \! \pmod{2}$, we have $f_1=f_{2^j+1}$.   
Noting that $|\cF^{(0)}_{2^j+1}|=2^{2^j}=|\S^{2^j}|$, 
we thus have $\cF^{(0)}_{2^j+1}=\{f^*f^*_1 : 
f^*=f^*_1f^*_2\ldots f^*_{2^j}\in \S^{2^j}\}$.
\endproof

\begin{lemma}
For $j\geq 0$, there is no periodic sequence $\x$ 
in $\cX_{2^j+1}$ whose period is $(2t+1)2^{j}$ for some $t\geq 0$. 
\label{small_per_lem}
\end{lemma} 
\mbox{}\\[-4ex]
\emph{Proof\/}: 
We deal with $j=0$ first. Note that $\cF_2^{(0)} = \{00,11\}$.
So, if $\cX_2$ has a periodic bi-infinite sequence 
$\w = (w_1 w_2 \ldots w_m)^\infty$ with an odd period $m$, 
then $00 \not\prec w_1w_2\ldots w_m$, 
$11 \not\prec w_1w_2\ldots w_m$, and $w_1 \neq w_m$. 
But there is no word $w_1w_2\ldots w_m \in \S^m$
that satisfies these conditions.

Now, consider $j \geq 1$.
Assume, to the contrary, that there exists a periodic sequence 
$\x=\ldots x_{-1}x_0x_1\ldots \in \cX_{2^j+1}$ 
whose period is $(2t+1)2^{j}$ for some $t\geq 0$. 
Then, $\x$ is of the form $(x_0x_1\ldots x_{(2t+1)2^j-1})^{\infty}$. 
Without loss of generality, we may assume that for every even integer $i$, 
$u\prec _i \x$ implies $u \not \in \cF^{(0)}_{2^j+1}$. Then, for each 
integer $m$, $x_{m2^j}x_{m2^j +1} \ldots x_{(m+1)2^j} \notin 
\cF^{(0)}_{2^j+1}$. So, by Lemma~\ref{forbidden_lem}, we have 
$x_{m2^j}\neq x_{(m+1)2^j}$. This implies that $x_0=x_{(2t)2^{j}}$ 
as $|\S|=2$. But then, 
$x_{(2t)2^j} \ldots x_{(2t+1)2^j-1}x_0 \in \cF^{(0)}_{2^j+1}$, 
which is a contradiction. 
\endproof \mbox{}\\[-20pt]

We are now in a position to prove Theorem~\ref{mult.thm}. \\[4pt]
\emph{Proof of Theorem~\ref{mult.thm}\/}: 
To prove the theorem, it is enough to show that for $j \geq 0$, 
the periods of periodic sequences in $\cX_{2^j+1}$ must be
multiples of $2^{j+1}$. It then follows, by Lemma~\ref{periodic_lem},
that the same also applies to periodic sequences in $\cX_k$, 
for $2^{j}+1 < k \leq 2^{j+1}$.

When $j=0$, the required statement clearly holds by Lemma~\ref{small_per_lem}. 
So, suppose that the statement is true for some $j \geq 0$, so that
periodic sequences in $\cX_{2^{j+1}}$ have only multiples of $2^{j+1}$ 
as periods. Therefore, by Lemma~\ref{periodic_lem}, periodic sequences 
in $\cX_{2^{j+1}+1}$ also can only have multiples of $2^{j+1}$ as periods. 
However, by Lemma~\ref{small_per_lem}, no periodic sequence in 
$\cX_{2^{j+1}+1}$ can have an odd multiple of $2^{j+1}$ as a period. 
Hence, all periodic sequences in $\cX_{2^{j+1}+1}$ have periods 
that are multiples of $2^{j+2}$. The theorem follows by induction.
\endproof \mbox{}\\[-20pt]

Theorem~\ref{mult.thm} shows that for $2^j+1 \leq k \leq 2^{j+1}$,
we have $T_{seq}^{(\cX_k)} \geq 2^{j+1}$. In fact, this holds with equality.
\\[-18pt]

\begin{corollary}
$T_{seq}^{(\cX_1)} = 1$, and for $k \geq 2$, if $j \geq 0$ is such that
$2^j+1 \leq k \leq 2^{j+1}$, then $T_{seq}^{(\cX_k)} = 2^{j+1}$.
\label{Tseq_cor}
\end{corollary}

\mbox{}\\[-4ex]
\emph{Proof\/}:
When $k=1$, $T_{seq}^{(\cX_1)}=1$ as $1^{\infty}\in \cX_1$. 
So let $k \geq 2$, and let $j \geq 0$ be such that 
$2^j+1\leq k\leq 2^{j+1}$. We only need to show that
$T_{seq}^{(\cX_k)} \leq 2^{j+1}$.
The bi-infinite sequence $\w=(0^{2^{j+1}-1}1)^{\infty}$ is in 
$\cX_{2^{j+1}}$ since, by Corollary~\ref{odd1_cor}, $\w$ contains no word in 
$\cF_{2^{j+1}}^{(0)}$ as a subword. Since $\w$ has period $2^{j+1}$,
by Lemma~\ref{periodic_lem}, $\w'= \psi^{2^{j+1}-k}(\w) \in \cX_k$ 
has period $2^{j+1}$ as well. Thus, $T_{seq}^{(\cX_k)}\leq 2^{j+1}$. 
\endproof 

Theorem~\ref{mult.thm} also implies the following corollary. \\[-18pt]

\begin{corollary}
$T_{graph}^{(\cX_k)} \geq T_{seq}^{(\cX_k)}$ holds when $1\leq k\leq 6$.  
\label{gra_seq_cor}
\end{corollary} 
\mbox{}\\[-4ex]
\emph{Proof\/}:
Since $\cX_1$ is proper, $T_{graph}^{(\cX_1)}\geq 2$ by Proposition~1 
in \cite{MS}. Thus, $T_{graph}^{(\cX_1)} > T_{seq}^{(\cX_1)} = 1$.

So, let $k\geq 2$ and suppose $2^j+1\leq k\leq 2^{j+1}$ for some $j\geq 0$. 
By Corollary~\ref{Tseq_cor}, we have $T_{seq}^{(\cX_k)} = 2^{j+1}$.
On the other hand, for any irreducible presentation 
$\cG$ of $\cX_k$, we have $\per(\cG)\geq 2^{j+1}$. Indeed, for each vertex 
$V$ in $\cG$, we have $\per(V)$ being a multiple of $2^{j+1}$;
otherwise we would have a contradiction of Theorem~\ref{mult.thm}.
Hence, $T_{graph}^{(\cX_k)}\geq 2^{j+1}=T_{seq}^{(\cX_k)}$ as required.   
\endproof \mbox{}\\[-20pt]

Corollary~\ref{Tseq_cor} shows that 
$T_{seq}^{(\cX_k)}$ grows arbitrarily large 
as $k \rightarrow \infty$, while $T_{desc}^{(\cX_k)} = 2$ for all $k$.
It also follows from Corollary~\ref{gra_seq_cor} 
that $T_{graph}^{(\cX_k)}$ is strictly larger than $T_{desc}^{(\cX_k)}$ 
when $3\leq k\leq 6$.
Equality can hold in Corollary~\ref{gra_seq_cor} --- for example, when $k=2$.
Indeed, $\cX_2$ is proper, and its MS presentation, $\cG_{\cX_2}$,
is irreducible, with $\per(\cG_{\cX_2})=2$, so that $T_{graph}^{(\cX_2)}=2$. 
From Corollary~\ref{Tseq_cor}, we also have $T_{seq}^{(\cX_2)}=2$.
Thus, $\cX_2$ is an example of a proper PFT $\cX$ in which 
$T_{seq}^{(\cX)}=T_{graph}^{(\cX)}=T_{desc}^{(\cX)}$ holds. 

Thus, to summarize, there appears to be no relationship between
the descriptive period of a PFT and its sequential period, as we have 
examples where each of these can be arbitrarily larger than the 
other. We have also found that, for a PFT $\cX$, $T_{graph}^{(\cX)}$
can be larger than $T_{desc}^{(\cX)}$. However, we believe
that the reverse cannot hold; in fact, we conjecture that 
$T_{desc}^{(\cX)}$ divides $T_{graph}^{(\cX)}$ for any PFT $\cX$.       

Finally, we note that we also have examples of proper PFT's $\cX$ where
$T_{seq}^{(\cX)}$ is arbitrarily larger than $T_{graph}^{(\cX)}$.
We omit the proof due to space constraints.

\begin{theorem}
Set $\S=\{0,1\}$ and $k\geq 2$, and let $\cP$ denote the set of all
periodic bi-infinite sequences over $\S$ with period $k!$. 
Consider the PFT $\cX = \cX_{\{\cF,2\}}$ with $\cF=(\cF^{(0)},\emptyset)$, 
such that $\cF^{(0)}=\{ w\in \Sigma^{2k!}\ : \exists \, \x \in \cP 
\text{ such that } w \prec \x\}$. The following 
statements hold: (a) $\cX$ is proper; (b) $\cX$ is irreducible; and 
(c) $T_{seq}^{(\cX)}\geq k+1$ and $T_{graph}^{(\cX)}=2$.
\label{seq>>graph_thm} 
\end{theorem}

\end{document}